\documentclass[prl,superscriptaddress,amsmath,amssymb,showpacs,noshowkes,a4paper,preprint]{revtex4}

\usepackage{graphicx}
\usepackage{dcolumn}
\usepackage{bm}
\usepackage[usenames]{color}
\usepackage{epstopdf}
\definecolor{mygray}{gray}{0.5}

\newcommand{\affmsc}{\affiliation{Laboratoire Mati\`ere et Syst\`emes
Complexes, Universit\'e Paris Diderot, CNRS - UMR 7057, B\^atiment
Condorcet, 10 rue Alice Domon et L\'eonie Duquet, 75013 Paris, France,
EU}}

\newcommand{\afflangevin}{\affiliation{Institut Langevin, ESPCI
ParisTech, CNRS - UMR 7587, 1 rue Jussieu, 75005 Paris Cedex 05, France,
EU}}

\begin{document}
\title{Chaos driven by interfering memory}

\author{S. Perrard}
\affmsc
\author{M. Labousse}
\afflangevin
\author{E. Fort}
 \thanks{Electronic address: \texttt{emmanuel.fort@espci.fr}; corresponding author}
\afflangevin
\author{Y. Couder}
\affmsc

\begin{abstract}
The transmission of information can couple two entities of very different nature, one of them serving as a memory for the other. Here we study the situation in which information is stored in a wave field and serves as a memory that pilots the dynamics of a particle. Such a system can be implemented by a bouncing drop generating surface waves sustained by a parametric forcing. The motion of the resulting ``walker" when confined in a harmonic potential well is generally disordered. Here we show that these trajectories correspond to chaotic regimes characterized by intermittent transitions between a discrete set of states. At any given time, the system is in one of these states characterized by a double quantization of size and angular momentum. A low dimensional intermittency determines their respective probabilities. They thus form an eigenstate basis of decomposition for what would be observed as a superposition of states if all measurements were intrusive.
\end{abstract}

\maketitle

One of the initial insights on the specificity of memory-based systems is due to E. Schr\"{o}dinger~\cite{Schrodinger_life_1944}. Discussing the need for stability of the biological transmission of information to progeny, he argued that a memory had to be encoded in a permanent structure of small size. Even though their observation in purely physical processes is still scarce, memory effects are no longer limited to biology; they appear in, \textit{e.g.} non-Markovian quantum effects~\cite{Chruscinski_PRA_2010,Kennes_PRL_2013}, crack propagations~\cite{Goldman_PRL_2010} looped neuronal networks~\cite{Neural_Memory_1997} or walking droplets~\cite{Fort_PNAS}. In all these systems, information is encoded and stored in various ways. The nature of the information repository defines the possible behaviors. Here we study the case of \textit{walkers} in which information is emitted and received by a localized object (a droplet) and stored in a spread-out surface wave. Because of interferences, the local object and its associated wave exhibit peculiar quantumlike duality~\cite{Walker_Nature,Couder_diffraction, Bush_PNAS,Fort_PNAS, Eddi_PRL_2012}. The present Letter is devoted to the emergence of chaos-induced statistical properties in this system. \\

The experiments are performed in a cell of diameter 13 cm containing a 6 mm deep layer of silicon oil of viscosity $\mu_L= 20 \times 10^{-3}$ Pa.s \cite{Perrard_natureC_2014} . It is oscillated vertically with an acceleration $\gamma=\gamma_0 \cos(2 \pi f_0 t)$ where $f_0=80$ Hz, and $\gamma_0$ is tunable. In this system, when $\gamma_0$ exceeds a threshold $\gamma_F = 4.5g$ (where g is gravity), a pattern of parametrically forced standing waves of frequency $f_0/2$, due to the Faraday instability, forms spontaneously. Our experiments are performed below this threshold. On the vibrated interface, a droplet of diameter $d\approx 0.7 mm$ of the same fluid can bounce indefinitely if $\gamma_0 > g$\footnote{The drop undergoes repeated collisions with the interface. It never merges with the liquid bath because the air film separating them has no time to drain. Each collision is partly elastic but most of the vertical momentum input is provided by an upward kick from the bath~\cite{Couder_2005}. The bouncing process has been investigated in detail by Mol\'{a}\v{c}ek and Bush~\cite{Molacek_JFM_1_2013}}. When $\gamma_0$ is larger than $3.5g$, the bouncing becomes sub-harmonic and the droplet acts as a local exciter of Faraday waves \cite{Couder_2005,Molacek_JFM_1_2013}. The droplet and the wave it generates are phase-locked. Correlatively, the drop starts moving at a velocity $V$ of the order of 9~mm/s. We call a \textit{walker} the resulting wave-particle association. Of particular relevance is the structure of the wave-field that drives the drop motion. At each impact, the drop excites a Bessel-like Faraday wave of period $T_F=2/f_0$ and wavelength $\lambda_F= 4.75$ mm, centered at the impact point. Since $\gamma_0<\gamma_F$, the waves are damped on a typical nondimensional time: $\mathrm{Me}=\tau/T_F$.  The global wave-field that pilots the drop is the linear superposition of all the waves generated by the successive impacts located along a memory length $S_{\mathrm{Me}}=V\tau/\lambda_F$ of the past trajectory. It thus contains in its interference pattern a path memory of the particle motion \cite{Fort_PNAS,Eddi_JFM_2011}. Since $\mathrm{Me}\approx \gamma_F/(\gamma_F-\gamma_0)$ \cite{Eddi_JFM_2011}, its value can be chosen by tuning $\gamma_0$ in the vicinity of $\gamma_F$. Previous works have shown that the memory has major effects on the drop motion whenever the walker is confined: in cavities~\cite{Harris_PRE_2013}, due to a Coriolis force~\cite{Harris_JFM_2014,Oza_JFM_2_2014}, or in a potential well \cite{Perrard_natureC_2014}. \\

\begin{figure}[h]
\begin{centering}
\includegraphics[width=7cm]{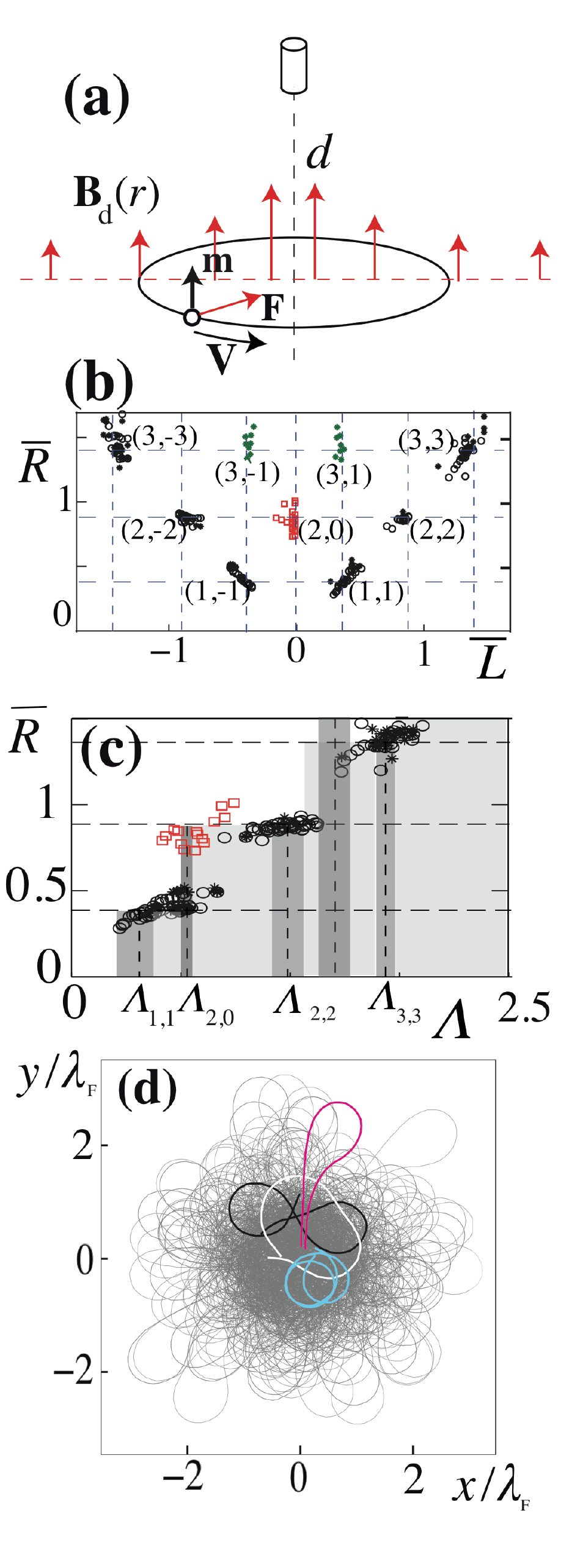}
\caption{\label{fig1} (a) Sketch of the experiment. The droplet loaded with ferrofluid is located in an axisymmetric spatially varying magnetic field $\mathbf{B}_{\mathrm{d}}(r)$ and thus trapped in a two-dimensional attractive harmonic potential well. (b) The eigenmodes defined by a plot of their mean nondimensional spatial extension $\bar R$ versus their mean nondimensional angular momentum $\bar L$. (c) When the control parameter $\Lambda$ changes. The stable modes $(n,m)$ are observed in narrow ranges of $\Lambda$ show in grey. (d) A highly intermittent trajectory of a drop of velocity $<V>=8.1$ mm.s$^{-1}$ at $\mathrm{Me}\approx 200$ for $\Lambda=0.83$ and $S_{\mathrm{Me}}/2\pi \Lambda=1.6$. The selected sections show the coexistence of orbits (1,$\pm$1), ovals (2,$\pm$2), lemniscates (2,0) and loops (3,$\pm$1).}
\end{centering}
\end{figure}

Here we investigate the latter situation obtained by applying a central force to the drop. The setup, schematized in Fig.~\ref{fig1}(a), is described in detail in Ref.~\cite{Perrard_natureC_2014}. The bouncing drop is loaded with a ferrofluid and polarized by a homogeneous magnetic field $\mathbf{B}_0$. It thus forms a magnetic dipole perpendicular to the bath surface. A magnet, placed on the cell's axis provides a second spatially varying magnetic field $\mathbf{B}_{\mathrm{d}}(r)$, where $r$ is the distance to the axis. The drop is thus trapped by a magnetic force: $\mathbf{F}(d) = -\kappa(d) \mathbf{r}$. The spring constant $\kappa$ can be tuned by changing the distance $d$ of the magnet to the liquid surface.  The walker confinement is controlled by the nondimensional half-width of the potential well $\Lambda = V \sqrt{m_W/\kappa}/\lambda_F$, where $m_W$ is the drop effective mass. The nature of the motion changes when the walker revisits regions where Faraday wave sources created in the past are still active. For orbital motions, this occurs when the memory length $S_{Me}=V \tau / \lambda_F$ is of the order of the nondimensional orbital perimeter $2 \pi \Lambda$. \\

In the high memory regime ($S_{\mathrm{Me}}>2\pi \Lambda$) the trapping leads to the appearance of states, as described previously \cite{Perrard_natureC_2014}. Each of them associates a stable periodic orbit with a specific global wave field. The orbits have different shapes (circles, ovals, lemniscates, trefoils, etc.) and two observables are needed to characterize them. Figure \ref{fig1}(b), adapted from ref.~\cite{Perrard_natureC_2014}, shows the mean nondimensional radius $\bar R = <\sqrt{R^2}/\lambda_F>$ as a function of the mean nondimensional angular momentum $\bar L=<L>/m_W \lambda_F <V>$. The experimental data are located at the nodes $(n,m)$ of a lattice as both observables can only take discrete values. They correspond for $\bar R$ to the successive zeros of the Bessel function $J_0(2 \pi r)$: $\left\lbrace r_1=0.37,r_2=0.87,r_3=1.87\right\rbrace $. For each given level $n$ the mean angular momentum $\bar L$ is also quantized: $\bar L_m$: $\{-r_n,-r_{n-2},...,0,...,r_{n-2},r_n\}$.
These states are only observed in narrow ranges of the well’s width $\Lambda_{n,m}^- < \Lambda <\Lambda_{n,m}^+$ centered around a set of discrete values $\Lambda_{n,m}$ [Fig.~\ref{fig1}()c]. \\
The present article deals with the complex trajectories observed when $\Lambda$ is tuned {\it outside} the stability ranges of the periodic orbits. Figure~\ref{fig1}(d) shows a typical example for $\mathrm{Me}\approx 200$ and $S_{\mathrm{Me}}/2\pi \Lambda = 1.6$. While the complexity increases with memory, it is remarkable that the regular orbits ($n$,$m$) still show up during short time intervals. In Fig.~\ref{fig1}(d) seven of them are present: circles (1,$\pm$1), lemniscates (2,0), ovals (2,$\pm$2), and loops (3,$\pm$1). \\
	
\begin{figure}[h]
\begin{centering}
\includegraphics[width=12cm]{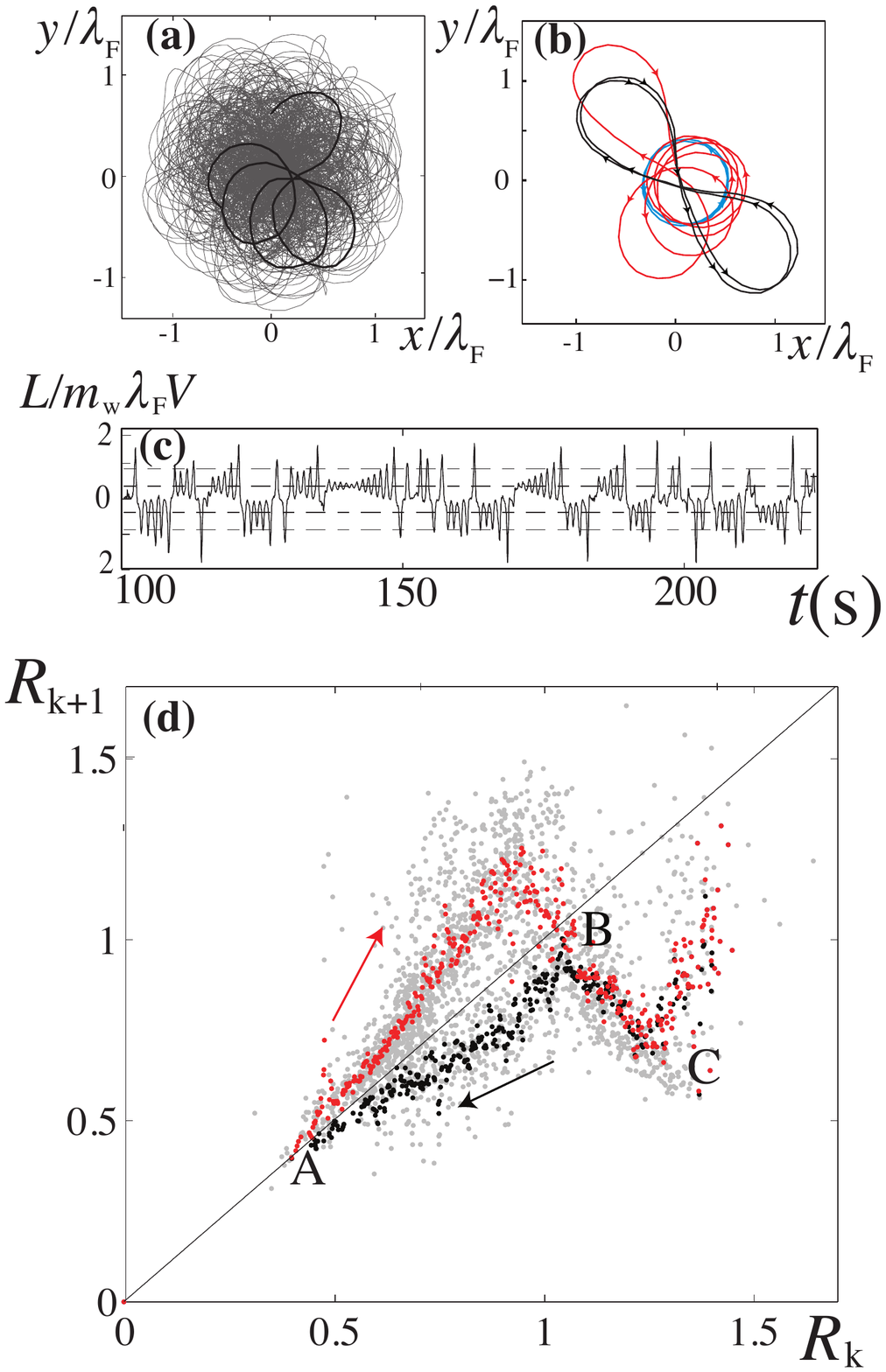}
\caption{\label{fig2} (a) A complex trajectory obtained for a walker of velocity $<V>=9.7$ mm.s$^{-1}$ and memory $\mathrm{Me}=63$, $S_{\mathrm{Me}}/2\pi \Lambda=1$ and for $\Lambda=0.49$, a value located in the range $\Lambda_{1,1}^+<\Lambda<\Lambda_{2,0}^-$. (b) A transition from a wobbling orbit (1,1) to a lemniscate (2,0). (c) A short sample of the time recording of the angular momentum $L$. The wobbling of increasing amplitude of a circular orbit (1,1) leads to a reversal (1,-1). These transitions are mediated by unstable lemniscates. (d) Map of first return of the local maximum $R_{k+1}$ as a function of the previous one $R_k$ for a recording lasting 40~min. The light grey points are all the iterates, the colored ones are averaged. The red dots correspond to increasing values of $R_k$ while the black ones are obtained for decreasing $R_k$.}
\end{centering}
\end{figure}

In order to put this coexistence of modes on a quantitative basis, we study the chaotic motion in the first two regions of instability ($\Lambda_{1,1}^+ < \Lambda <\Lambda_{2,0}^-$ and $\Lambda_{2,0}^+ < \Lambda <\Lambda_{2,2}^-$), for intermediate values of the memory for which $S_{\mathrm{Me}}/2 \pi \Lambda$ is close to 1. Figure~\ref{fig2}(a) shows an example of a complex trajectory obtained for $\Lambda$=0.49 at a memory ($\mathrm{Me}\approx$ 63 and $S_{\mathrm{Me}}/2\pi \Lambda \approx 1$). Only three unstable states, the small orbits (1,$\pm$1) and the lemniscate (2,0) coexist. Figure~\ref{fig2}(b) shows a typical scenario of circle destabilization. It originates in a mismatch of the classical orbiting radius (due to the central force) and the orbiting radius induced by the wave field. The transition from a circle to a lemniscate, occurs when the wobbling brings the trajectory close to the center. A topological change then leads to a lemniscate. This multi-looped motion appears unstable and mediates a return to orbiting motion with a possible flip of the angular momentum~\cite{Berhanu_EPL_2007,Petrelis_PRL_2009}~\footnote{The sign reversal of $\bar L$ is mediated by the existence of an intermediate unstable state of different symmetry. This phenomenon was discussed in the context of the magnetic dipole reversals in the earth dynamo effect~\cite{Petrelis_PRL_2009}.}. Figure~\ref{fig2}(c) shows the time recording of $L$ associated to the trajectory of figure~\ref{fig2}a. The transitions are typical of low-dimensional chaos in dissipative system \cite{Eckmann_1981,Pomeau_1980, Berge_1984}.

	The multistability can be characterized using a map of first return relating the nondimensional distance $R$ to the center at time $t+T_I$ to its value at time $t$. The discretization is obtained by considering the evolution of the successive maxima. The time interval $T_I$ is then self-determined. The resulting iterative map of $R_{\mathrm{max}}(k+1)$ as a function of $R_{\mathrm{max}}(k)$ is shown in figure~\ref{fig2}(d). The dynamics is described by the application of first return. The two fixed points A and B correspond to circles and lemniscates respectively. Here, the tuning value of $\Lambda$ sets the system in a regime where both these attractors are unstable. Starting from A, the wobbling grows corresponding to a drift from A to B along the upper branch. In the neighborhood of B, the motion is a lemniscate. After a few loops, its instability triggers a return to A. The route back depends on the memory. For $\mathrm{Me}\approx 40$ with $S_{\mathrm{Me}}/2\pi \Lambda \approx 0.8$, the iterative points drift directly from C to the upper branch in the vicinity of A. For a slightly larger memory $\mathrm{Me}=60$ with $S_{\mathrm{Me}}/2\pi \Lambda \approx 1$, the transition is mediated by an orbital motion of decreasing wobbling amplitude. This is observed as a new branch in the map. The first return application becomes multi-valued and the dynamics can no longer be described in a deterministic manner by a one-dimensional map: another dimension has to be added. This new degree of freedom originates in the role of information stored in the wave field. For each periodic orbit, the global wave-field can be decomposed in a Bessel function basis $J_k(2\pi r)$ centered at the magnet axis. For instance, a circular state ($m=\pm n$) is known to inhibit the $J_0$ mode and a lemniscate to minimize the $J_2$ mode~\cite{Perrard_natureC_2014}. The two branches of the iterative map could be distinguished by, e.g., the amplitude of the $J_0$ mode. This iterative map shows that the quantized lemniscates and circular orbits remain fixed points of the time evolution of $R$ but both unstable. The instability increases with memory. This is a non-Markovian process : the past contributes to the instability of the present. As a result the typical time between flips decreases with increasing memory, an effect beyond the scope of the present Letter.

\begin{figure}[h]
\begin{centering}
\includegraphics[width=12cm]{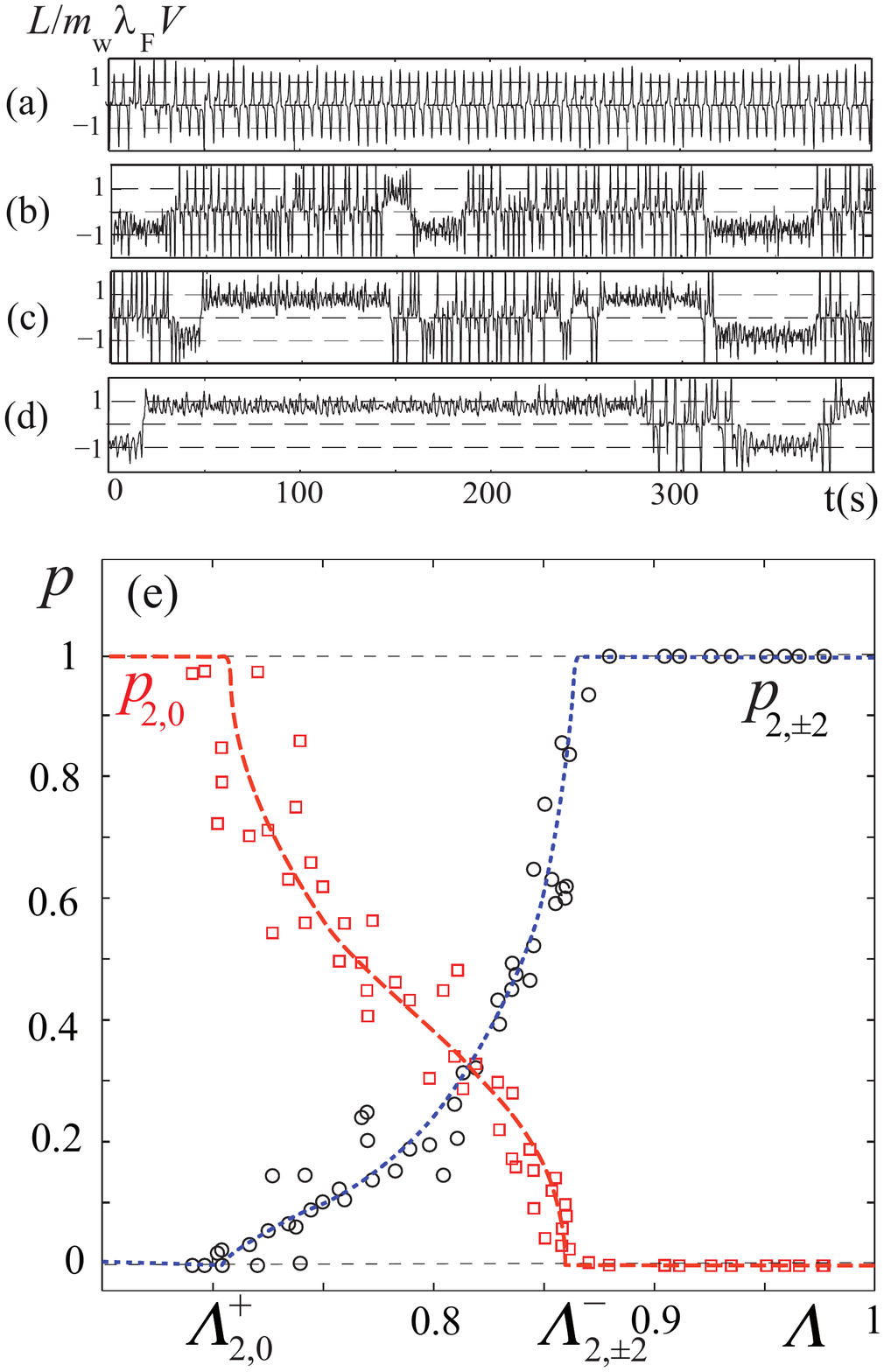}
\caption{\label{fig3} Analysis of the evolution of the multi-stability obtained for increasing values of the control parameter in a range where the three eigenstates corresponding to $n=2$ are lemniscates (2,0) and oval shaped orbits (2,$\pm$2).  (a)-(d) Four time recordings of the normalized angular momentum $L$  obtained for a walker of velocity $<V>=9.7$ mm.s$^{-1}$ for $\mathrm{Me}=60$ are shown corresponding to $\Lambda=0.74$, $\Lambda=0.807$, $\Lambda=0.828$, $\Lambda=0.875$, respectively. The typical intermittency time can be large, e.g., $\sim$150 orbital periods in (d). (e) Evolution of the probability $p_{2,0}$ and $p_{2,\pm 2}$ of being in lemniscate (square) and  oval shaped states (o) as a function of $\Lambda$ obtained with 12 different drops for $\mathrm{Me} = 60 \pm 10$. The sum of
the two probabilities $p_{2,\pm 2} + p_{2,0}$ is close to 1.}
\end{centering}
\end{figure}

Particularly interesting is the intermittency between two levels having the same spatial extension. Figure~\ref{fig3} presents the multistability observed in the range $\Lambda^{+}_{2,0} < \Lambda < \Lambda^{-}_{2,\pm 2}$. Four time recordings of the nondimensional angular momentum for increasing values of $\Lambda$ are given in Fig.~\ref{fig3}(a)-\ref{fig3}(d). The signal is not erratic, but composed of three types of domains. In some of them $\bar L \approx \pm0.87$ while in the third one, $L$ oscillates rapidly so that $\bar L \approx 0$. The corresponding trajectory fragments are oval orbits (2,$\pm$2) and lemniscates (2,0), respectively. The system thus spends time in a given eigenstate before undergoing an abrupt transition to another. Such intermittency is observed for all values of $\Lambda$ in the range separating the pure lemniscate at $\Lambda^{+}_{2,0}$ from the pure circle at $\Lambda^{-}_{2,\pm 2}$. Using long recordings of duration $T_R\simeq $ 1500 s, we measure the probabilities $p_{2,+2}$, $p_{2,-2}$ and $p_{2,0}$ of being in each state. We find $p_{2,+2}$ and $p_{2,-2}$ to be equal. As shown in Fig.~\ref{fig3}(e), their sum $p_{2,\pm 2}$ increases while $p_{2,0}$ decreases when $\Lambda$ goes from $\Lambda_{2,0}^+$ to $\Lambda_{2,\pm 2}^-$ . The sum $p_{2,\pm 2} + p_{2,0}$ is close to but slightly smaller than 1 because the angular momentum is ill-defined during the mode switching.

	In our previous work~\cite{Perrard_natureC_2014}, we had shown that the confinement of walkers led to stable orbital motions only if both the orbit size and the mean angular momentum satisfy quantization conditions. This is a nonquantum quantization: our system has no relation with the Planck constant. As discussed in ref.~\cite{Fort_PNAS}, an analogy appears by considering that the Faraday wavelength here plays a role similar to the de Broglie wavelength in quantum mechanics. Here, we have demonstrated how the chaotic motion is characterized by transitions between these periodic orbits. Even in a complex situation, the walker, at each given time, is in one of the possible discrete modes. At higher memory [see Fig.~\ref{fig1}(d)] the phenomena are the same but the number of modes involved in the decomposition increases. This justifies the use of the term eigenstate in Ref.~\cite{Perrard_natureC_2014} for the states $(n,m)$: they form a basis of decomposition on which complex trajectories can be decomposed. Only a perfect tuning of $\Lambda$ permits the preparation a "pure" state for which any time evolution appears forbidden. Finally, we can note that we have in this system the opportunity of performing a continuous nonintrusive observation. It is interesting to consider a \textit{gedanken} situation in which, in this system, we could only have a single glimpse of duration $T_I$, during which an intrusive measurement would be done. The glimpse would lead to the observation of one of the eigenstates only. With many realizations, a probability of each result would emerge. A superposition of states would then be the best reachable description and the probabilities would appear intrinsic. 

\acknowledgements{The author are grateful to S. Fauve, O. Giraud, A. Libchaber and F. P\'etr\'elis for useful discussions, and to L. Rh\'ea, D. Charalampous and A. Lantheaume for technical assistance. We thank AXA Research Fund and the French National Research Agency (Labex Wifi, Project Freeflow) for financial support.}

\bibliography{bibliogouttes}
\newpage

\end{document}